\documentclass{PoS}

\def\be{\begin{equation}}
\def\en{\end{equation}}
\def\mt{ }

\title{The variable Crab Nebula}

\ShortTitle{The variable Crab Nebula}

\author{\speaker{Marco Tavani}\thanks{Report presented on behalf of the AGILE Team.}\\
        INAF-IASF-Rome and University of Tor Vergata, Rome (Italy)\\
        E-mail: \email{tavani@iasf-roma.inaf.it}}

\abstract{ The remarkable Crab Nebula is powered by an energetic
pulsar
whose relativistic wind
interacts with the inner parts of the Supernova Remnant SN1054.
 Despite low-intensity optical and X-ray variations in the inner Nebula, the Crab
has been considered until now substantially stable at  X-ray and
gamma-ray energies.  This paradigm has been shattered by the AGILE
discovery in September 2010 of a very intense transient gamma-ray
flare of nebular origin.
 For the first time,
 the Crab Nebula was "caught in the act" of
accelerating particles up to $10^{15}$ eV within the shortest
timescale ever observed in a cosmic nebula (1 day or less).
Emission between 50 MeV and a few GeV was detected with a quite
hard spectrum within a short timescale. Additional analysis and
recent Crab Nebula data lead to identify a total of four major
flaring gamma-ray episodes detected by AGILE and \textit{Fermi}
during the period mid-2007/mid-2011. These observations challenge
emission models of the pulsar wind interaction and particle
acceleration processes. Indeed, the discovery of fast and
efficient gamma-ray transient emission from the Crab  leads to
substantially revise current models of particle acceleration. }

\FullConference{25th Texas Symposium on Relativistic Astrophysics - TEXAS 2010\\
        December 06-10, 2010\\
        Heidelberg, Germany}

\begin{document}

\section{The context}

The Crab Nebula is probably the most studied source in high-energy
astrophysics. It is at the center of the SN1054 supernova remnant,
and consists of a rotationally-powered pulsar interacting with a
surrounding nebula through a very powerful relativistic
wave/particle wind (e.g., refs. \cite{rees,hester1}).
Many observations and theoretical modelling were devoted to the
study of this remarkable system. A powerful pulsar located at its
center (of spindown luminosity $\textit{L$_{PSR}$} $=
5$\cdot$10$^{38}$ erg s$^{-1}$, and spin period $\textit{P}$ = 33
ms) is energizing the whole system with its wave/particle output.
The Nebula and its pulsar emit unpulsed and pulsed emission,
respectively, in a broad-band spectrum ranging from radio to TeV
energies.
 The Crab  is then an ideal laboratory to study very
 efficient particle acceleration subject to the different physical conditions
of the following sites: (\textbf{S1}) the pulsar magnetosphere,
(\textbf{S2}) the relativistic pulsar wind, (\textbf{S3}) the
relativistic termination shocks, (\textbf{S4}) MHD and/or plasma
flow instabilities, (\textbf{S5}) large-scale nebular
interactions.

The inner Nebula shows distinctive and variable optical and X-ray
brightness enhancements (``wisps'', ``knots'', and the ``anvil''
aligned with the pulsar ``jet'')
\cite{scargle,hester1,hester2,hester3,weisskopf}. Time variations
of these features (weeks, months) have been attributed to local
enhancements of the synchrotron emission related to instabilities
and/or shocks in the pulsar wind outflow. These variations are
indicative of changing acceleration tied to the physical processes
originating especially in sites S3 and S4. However, this variable
local activity in the inner Nebula, remarkable as it is
\cite{hester2,hester3}, has been known to produce only small
percentage variations of the total Crab radio, optical, X-ray and
gamma-ray energy fluxes. It is  not surprising that, when averaged
over the whole inner region (several arcminute across), the Crab
Nebula has been considered essentially stable, and used as a
"standard candle" in high-energy astrophysics.

Synchrotron and inverse Compton modelling  of the Crab Nebula
emission produces a reasonable picture of its {\it average}
properties (e.g.,
\cite{rees,kennel,dejager1,dejager2,atoyan,arons,meyer}]. The MHD
pulsar wind interacts with its environment through a sequence of
"shocks" or dissipation features localized at distances larger
than a few times $10^{17}$cm. Particle acceleration processes
(mostly in the pulsar wind and termination shock regions) produce
at least  two main populations of accelerated electrons/positrons
that explain the radio/optical  and the X-ray/gamma-ray emissions,
respectively \cite{dejager2,atoyan,meyer}. The optical/radio and
the X-ray continuum and gamma-rays up to $\sim 50-100 \, $MeV
energies are modelled by synchrotron radiation of accelerated
particles in an average nebular magnetic field $\bar{B} = 200 \,
\mu$G \cite{hester2,dejager2,atoyan,meyer}. Radiation from GeV to
TeV energies is modelled as the sum of different contributions of
inverse Compton components of electrons/positrons scattering CMB
and nebular soft photons \cite{dejager1,dejager2,atoyan,meyer}.

Acceleration at the pulsar wind termination shock regions S3 is
crucial for our purposes. The Kennel-Coroniti analysis of the
Nebula indicating a low-sigma pulsar wind \cite{kennel} favors
"strong" shocks at S3. Different acceleration models have been
proposed, including diffusive processes (e.g.,
\cite{blandford,dejager1,dejager2,atoyan}), shock-drift
acceleration (e.g., Begelman \& Kirk 1990) or ion-mediated
acceleration (e.g., \cite{arons,spitkovsky}). In general,
diffusive acceleration models imply acceleration rates of order of
the relativistic electron cyclotron frequency $\omega_B/\gamma$ (
with $\omega_B = e \, B / m_e \, c$, and $\gamma$ the electron's
Lorentz factor), that is $\tau_{acc} \simeq \alpha' \,
\omega_B/\gamma = c/R_L $ with $\alpha' \leq 1$ an efficiency
parameter and $R_L$ the Larmor radius (e.g.,
\cite{dejager1,dejager2,atoyan}). If we assume: (i) a Doppler
factor $\delta = \Gamma^{-1} \, (1 - \beta \cos(\theta))^{-1} = 1$
(with $\theta$ the emission angle with respect to the line of
sight ), (ii) $\alpha' \simeq 1$, (iii) the equality between the
accelerating electric field ($E$) and the magnetic field at the
acceleration site, and (iv) synchrotron cooling in the co-spatial
magnetic field, we obtain a most quoted constraint for the maximum
radiated photon energy $\epsilon_{\gamma,max}$
\cite{dejager1,dejager2}
\be \epsilon_{\gamma,max} \simeq \, \frac{m_e \, c^2}{\alpha}
\simeq 50 \; \rm MeV \label{eq-1} \en
with $\alpha = e^2/\hbar \, c$ the fine structure constant, $c$
the speed of light, and $m_e$ the electron's mass. Eq. \ref{eq-1}
applies in a natural way to {\mt particles "slowly" accelerated},
e.g., by diffusive processes, and $E_{\gamma,max}$ turns out to be
independent of the local magnetic field. The quasi-exponential
cutoff shown by the average Crab Nebula gamma-ray spectrum in the
10 MeV - 10 GeV range supports this idealization
\cite{dejager1,dejager2}. However, the local equality between the
electric and magnetic fields (expected in MHD models and/or
electromagnetic turbulent wave heating) should not be imposed.
Considering also the possible effects of a non-trivial Doppler
factor $\delta \neq 1$ and acceleration efficiency $\alpha' \neq
1$, the most general expression for the maximum
synchrotron-emitted photon energy is\footnote{See also ref.
\cite{aharonian2}.}

\be \epsilon_{\gamma,max} \simeq \frac{9}{4} \, \left( \frac{E}{B}
\right) \frac{m_e \, c^2}{\alpha}  \, \left( \frac{\delta \,
\alpha'}{\langle \sin(\theta') \rangle} \right) \simeq (150 \;
{\rm MeV}) \, \left( \frac{E}{B} \right) \, \left( \frac{\delta \,
\alpha'}{\langle \sin(\theta') \rangle} \right) \label{eq-2} \en

\noindent where $\langle \sin(\theta') \rangle $ takes into
account the effects of the average pitch angle $\theta'$. We
expect the combination $\delta \, \alpha' /\langle \sin(\theta')
\rangle $ to be of order unity as confirmed by the average steady
emission of the Crab Nebula showing the "synchrotron burn-off"
above 10 MeV \cite{dejager2}. Expectations from MHD-based models
of emission are clear, and well represented by Eq.~\ref{eq-2}.
This theoretical framework is now challenged by the AGILE
discovery of the strong gamma-ray flare from the Crab Nebula in
September 2010 \cite{tavani1,tavani3}, as we describe below.

\section{The discovery}

The AGILE satellite \cite{tavani2}  observed the Crab Nebula
several times both in  pointing mode  from mid-2007/mid-2009, and
in  spinning mode  starting in November 2009. The AGILE instrument
monitors cosmic sources in the energy ranges 50 MeV - 10 GeV
(hereafter, GeV gamma-rays) and 18 - 60 keV with good sensitivity
and angular resolution. With the exception of occasional flaring
episodes
(see below), AGILE detects an average (pulsar +nebula) flux value
$F_{ave} = (2.2 \pm 0.1)\times 10^{-6} \, \rm ph \, cm^{-2} \,
s^{-1} $ in the range 100 MeV - 5 GeV \cite{pittori}, for an
average photon index  $\alpha = 2.13 \pm 0.07$.

During routine monitoring in spinning mode in September 2010, a
strong and unexpected gamma-ray flare from the direction of the
Crab Nebula was discovered by AGILE above 100 MeV and immediately
reported to the community \cite{tavani1}. The flare reached its
peak during 19-21 September 2010 with a 2-day flux of $F_{g,p1}=
(7.2 \pm  1.4)\times 10^{-6} \, \rm ph \, cm^{-2} \, s^{-1} $
($\alpha = 2.03 \pm 0.18$) for a 4.8 s.d. detection above the
average flux. The flux subsequently decayed within 2-3 days to
normal average values (Fig. 1) \cite{tavani3}. This flare was
confirmed within 1 day by \textit{Fermi}-LAT \cite{buehler}, and
different groups obtained multifrequency data during the  days
immediately following the flare (see ref. \cite{bernardini} for a
summary of follow-up observations). Recognizing the importance of
this event was facilitated by a previous AGILE detection of the
Crab Nebula with similar characteristics. Indeed, AGILE detected
another remarkable flare from the Crab in October, 2007 (see also
\cite{pittori}). This flare lasted $\sim$2 weeks and showed  an
interesting time sub-structure (Fig.~2, top panel). The peak flux
was reached on Oct. 7, 2007, and the 1-day integration value was
$F_{g,p2} =(8.9 \pm 1.1) \times 10^{-6} \rm \, ph \, cm^{-2} \,
s^{-1}$ ($\alpha = 2.05 \pm 0.13)$ for a 6.2 s.d. detection above
the standard flux.

 \begin{figure}[ht!]
\begin{center}
 \includegraphics[width=11cm]{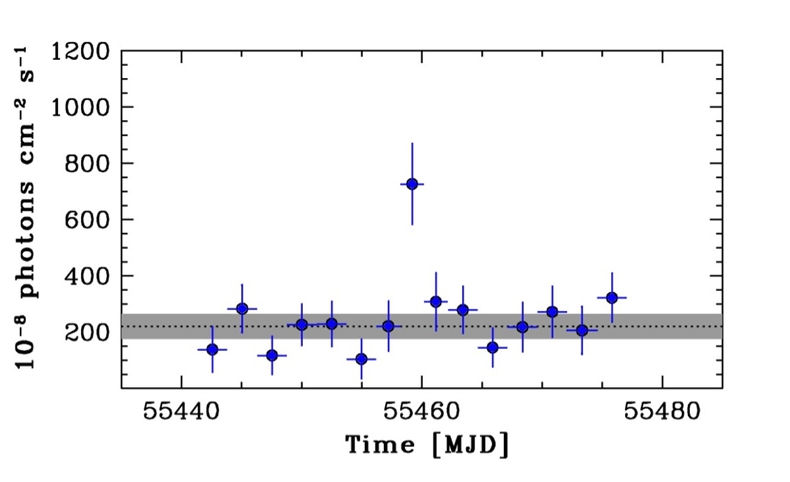}
\caption{The AGILE gamma-ray lightcurve of the Crab pulsar $+$
nebula above 100 MeV showing the September 2010 flare
\cite{tavani3}. The band marked in grey shows the average flux
within 3 standard deviations.}
 \label{fig-1}
 \end{center}
 \end{figure}

\begin{figure}[ht!]
\begin{center}
\vspace*{-0.5cm}
\includegraphics[width=10.cm]{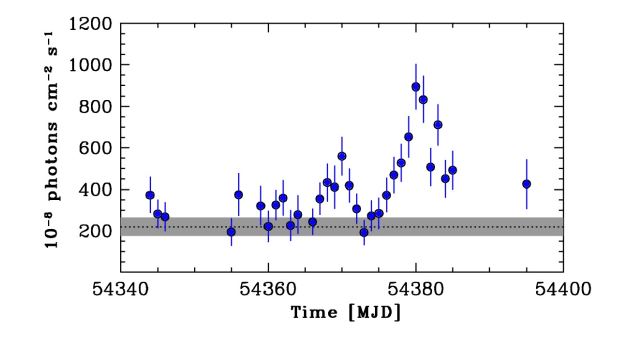}
\vspace*{-.01cm}
\includegraphics[width=8.cm]{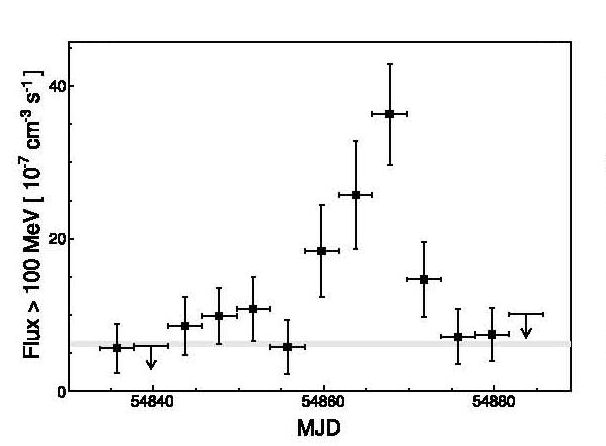}
\vspace*{-0.1cm}
\includegraphics[width=10cm]{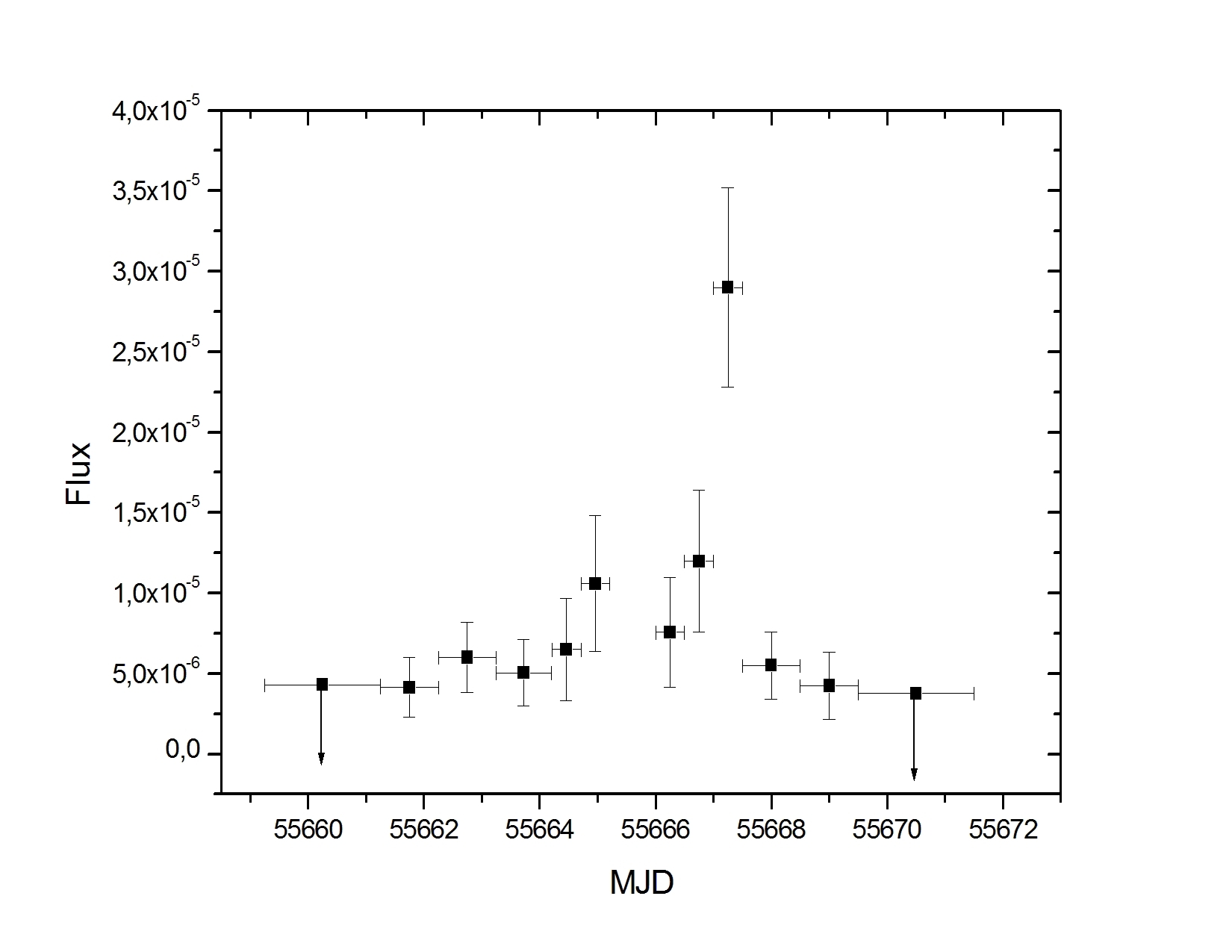}
\caption{\textit{Upper panel:} the gamma-ray lightcurve of the
Crab pulsar $+$ nebula above 100 MeV showing the October, 2007
flare detected by AGILE \cite{tavani3}.  The band marked in grey
shows the average flux within 3 s.d. \textit{Middle panel:} the
gamma-ray lightcurve of the Crab nebula above 100 MeV showing the
February, 2009 flare detected by \textit{Fermi}-LAT \cite{abdo2}.
The band marked in grey shows the average Nebula flux in the
\textit{Fermi}-LAT energy range. \textit{Lower panel:} gamma-ray
lightcurve of the Crab pulsar $+$ nebula above 100 MeV showing the
October, 2007 flare detected by AGILE \cite{striani}. }
 \label{fig-2}
 \end{center}
 \end{figure}

The pulsed emission from the Crab pulsar is characterized very
well from radio, optical, X-ray up to gamma-ray energies
\cite{fierro,abdo,pellizzoni}. For all major flares and especially
for the October 2007 and September 2010 events there was no sign
of variation of the pulsar signal before, during and after these
flares (e.g., supplemental information in \cite{tavani3}). For the
Sept. 2010 event the issue of search for possible variation of the
pulsed emission was thoroughly studied at radio \cite{espinoza},
X-ray \cite{tavani3} and gamma-ray \cite{hays,abdo2,tavani3}
energies.
 Thus, we attribute the major gamma-ray flares from the Crab
to transient emission  originating from relativistic particles
accelerated in the inner Nebula, or in regions decoupled from the
rotating magnetosphere producing the pulsed
emission\footnote{There is no evidence for an additional
transient source (flaring blazar or Galactic source) near the Crab
Nebula for the Sept. 2010 event or for other major gamma-ray
flares. A typical Crab Nebula gamma-ray error box radius for long
integrations is $r = 0.1^o$ or less, for both AGILE and
\textit{Fermi}-LAT (e.g., \cite{abdo2}). In particular, for the
September 2010 event no change in the hard X-ray flux was detected
by Integral and Swift/BAT \cite{ferrigno,markwardt}, and no other
X-ray source was detected by \textit{Swift}/XRT  about 3 days
after the event (0.7 mCrab upper limit for a source in an annular
region of size between 1 and 2 arcminutes, and 0.3 mCrab upper
limit in an annular region of radius between 3 and 10 arcminutes,
\cite{heinke}). The peculiar flaring gamma-ray spectra and the
very high intensity reached in 2010 and 2011 make the hypothesis
of a background transient source even more improbable. Depending
on assumptions, the probability of random occurrence of a peculiar
transient source in the Crab Nebula gamma-ray error box is in the
range $10^{-4} - 10^{-6}$ \cite{abdo,tavani3}. }.

\section{Major gamma-ray flares}


 \begin{figure} 
\begin{center}
\vspace*{-.5cm}
\includegraphics[width=12.cm]{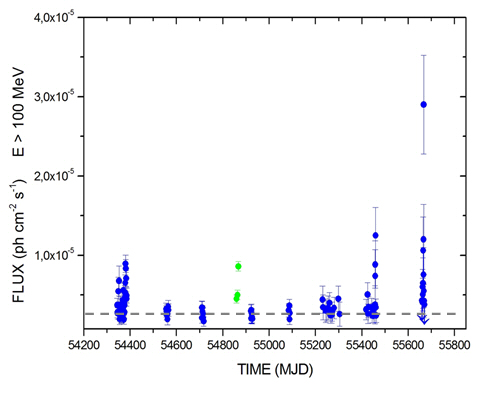}
\vspace*{-0.2cm}
 \caption{The AGILE gamma-ray lightcurve of the
Crab Nebula (pulsar plus nebula) above 100 MeV monitored from
mid-2007  until May 2011. The data points marked in green show the
gamma-ray flare reported by \textit{Fermi}-LAT \cite{abdo2}. The
dashed line marks the average flux in the AGILE-GRID energy
range.}
 \label{fig-3}
 \end{center}
 \end{figure}


Four major gamma-ray flaring episodes from the Crab Nebula have
been reported during the period mid-2007/mid-2011 by AGILE
\cite{tavani3,tavani4,striani4,striani} and \textit{Fermi}-LAT
\cite{abdo2,buehler4,hays4} (see
 Table~1).
%
The event characteristics and total radiated energies  differ from
one event to the other.
The first two events (in 2007 and 2009) are relatively "slow" in
their risetime (a few days) and in the overall temporal evolution.
Their duration is at least 2 weeks and the total gamma-ray
radiated energy is  $E_{\gamma} \sim 10^{42}$~ergs. The third
event (Sept. 2010) detected by both AGILE and \textit{Fermi}-LAT
developed on a faster timescale. Its risetime is less than 1 day
\cite{tavani3}, and the (conservative) lightcurves published by
the AGILE and \textit{Fermi}-LAT teams do not show significant
emission beyond 4 days. However, there is evidence that the
temporal structure of the Sept. 2010 event may be  non-trivial as
suggested by an analysis of \textit{Fermi}-LAT data \cite{balbo}
that has been recently confirmed by AGILE \cite{striani}. The
Sept. 2010 event shows variability on a 12-hr timescale with a
strong variation of the gamma-ray flux. In this case, the  total
gamma-ray radiated energy is  $E_{\gamma} \sim 10^{41}$~ergs.

Even more dramatic time variability has been recently detected in
the most recent and remarkable Crab flaring event of April, 2011
\cite{buehler4,tavani4,hays4,striani4}. In this case, variability
at the level of a few hours (and possibly less) has been
determined \cite{striani,hays4}. The April 2011 event can be
called a  Crab gamma-ray "super-flare", with the flux above 100
MeV reaching the record-high peak of $F = (30 \pm 6) \times
10^{-6} \rm \, ph \, cm^{-2} \, s^{-1}$ on a 12-hour timescale
\cite{striani}.  This event lasts about 1 week and the  total
energy emitted above 100 MeV is  $E_{\gamma} \sim 10^{42}$~ergs.

Fig.~\ref{fig-3} shows the result of the AGILE long baseline
gamma-ray monitoring of the Crab Nebula from mid-2007 until May,
2011 together with the Feb. 2009 flare detected by
\textit{Fermi}-LAT. It is interesting to note that all three
events appear to be characterized by a "precursor" (lasting a few
days) that anticipates a major peak flare.
 The Crab Nebula is producing major gamma-ray flares
with an apparent rate of (1-2)/year. Occasionally, the level of
gamma-ray emission is surprisingly high, reaching the level of the
brightest gamma-ray blazars. In April, 2011 the Crab Nebula became
for a few days the most powerful source in the  GeV gamma-ray sky,
exceeding the Vela pulsar flux by a factor of almost three !

\begin{table}[h!]
\begin{center}
{\bf Table 1: Major gamma-ray flares of the Crab Nebula} \vskip
.1in
\begin{tabular}{|c|c|c|c|c|}
  \hline
  \textbf{Flare date} & \textbf{Duration} & \textbf{Peak flux (*)}  & \textbf{Instrument} & \textbf{Ref.} \\
                      &                   & (pulsar + nebula)  & & \\
  \hline
  October 2007 & $\sim$ 15 days & $ 9 \times 10^{-6} \; \rm ph \, cm^{-2} \,
s^{-1} $ & AGILE & \cite{tavani3} \\
  February 2009 & $\sim$ 15 days & $ 7 \times 10^{-6} \; \rm ph \, cm^{-2} \,
s^{-1} $  & \textit{Fermi}-LAT & \cite{abdo2}  \\
  September 2010 & $\sim$ 4 days &  $ 7 \times 10^{-6} \; \rm ph \, cm^{-2} \,
s^{-1} $ & AGILE, \textit{Fermi}-LAT &  \cite{tavani3,abdo2} \\
April  2011 & $\sim$ 10 days &  $ 30 \times 10^{-6} \; \rm ph \,
cm^{-2}
\, s^{-1} $ & AGILE, \textit{Fermi}-LAT & \cite{buehler4,tavani4,hays4,striani4} \\
  \hline
\end{tabular}
\end{center}
* Average peak fluxes obtained for different integration times: 1
day (October 2007), 4 days (February 2009), 2 days (September
2010), 12 hours (April 2011).
\end{table}


\section{Modelling the flaring gamma-ray spectrum}


\begin{figure}
\begin{center}
\vspace*{-0.5cm}
\includegraphics[width=11cm]{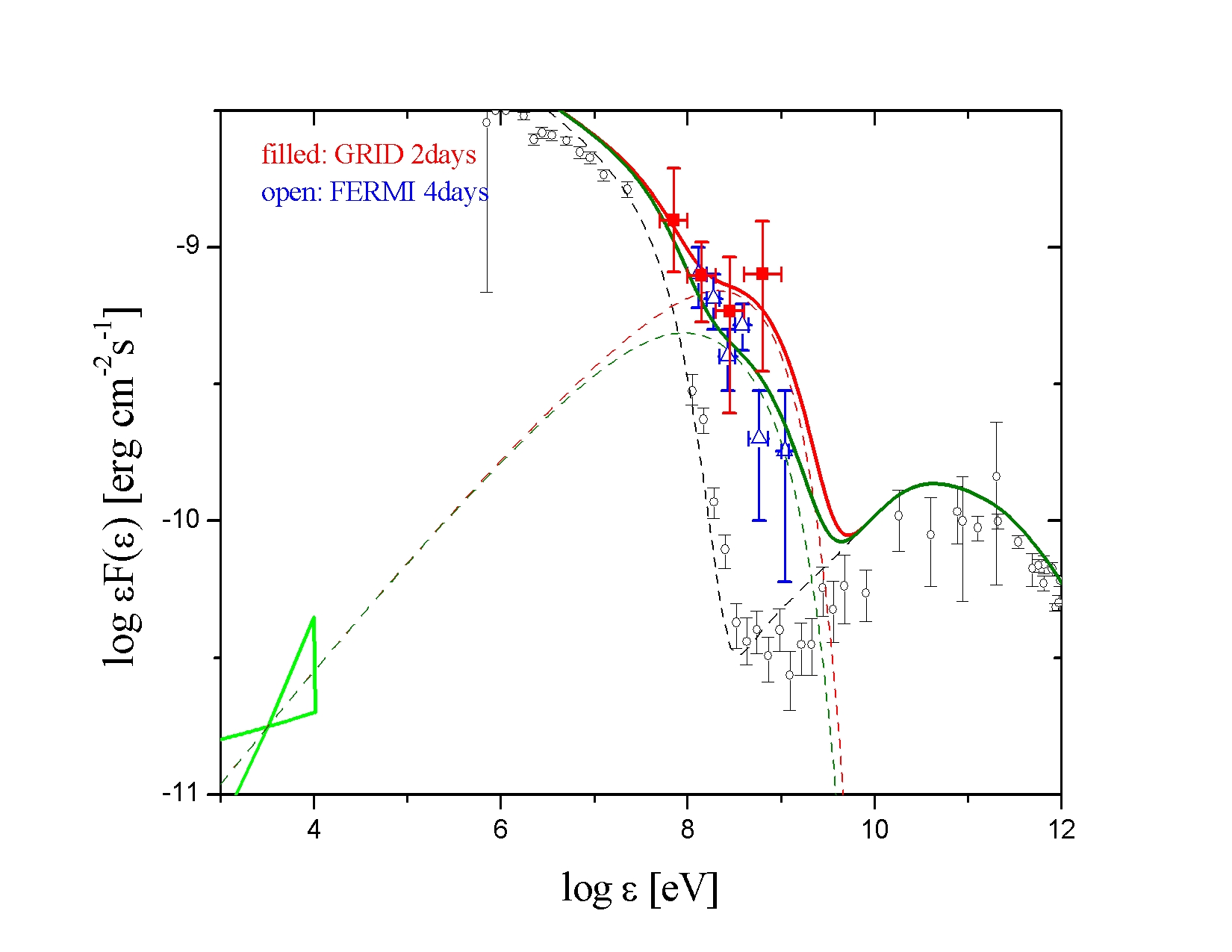}
\vspace*{-0.2cm}
\caption{Gamma-ray (pulsar subtracted) spectra of
the Crab Nebula September 2010 flare. Red data: AGILE spectrum
integrated over a 2-day interval \cite{tavani3}. Blue data:
\textit{Fermi}-LAT spectrum integrated over a 4-day interval
\cite{abdo2}. Note the different integration times. Black data
points show the standard Crab Nebula spectrum \cite{meyer}. The
area marked in green shows the X-ray emission spectrum detected by
Chandra about 1 week after the Sept. 2010 flare from "Source A" in
the inner Nebula as reported in ref. \cite{tavani3}. }
 \label{fig-4}
 \end{center}
 \end{figure}

\begin{figure}
\begin{center}
\includegraphics[width=14cm]{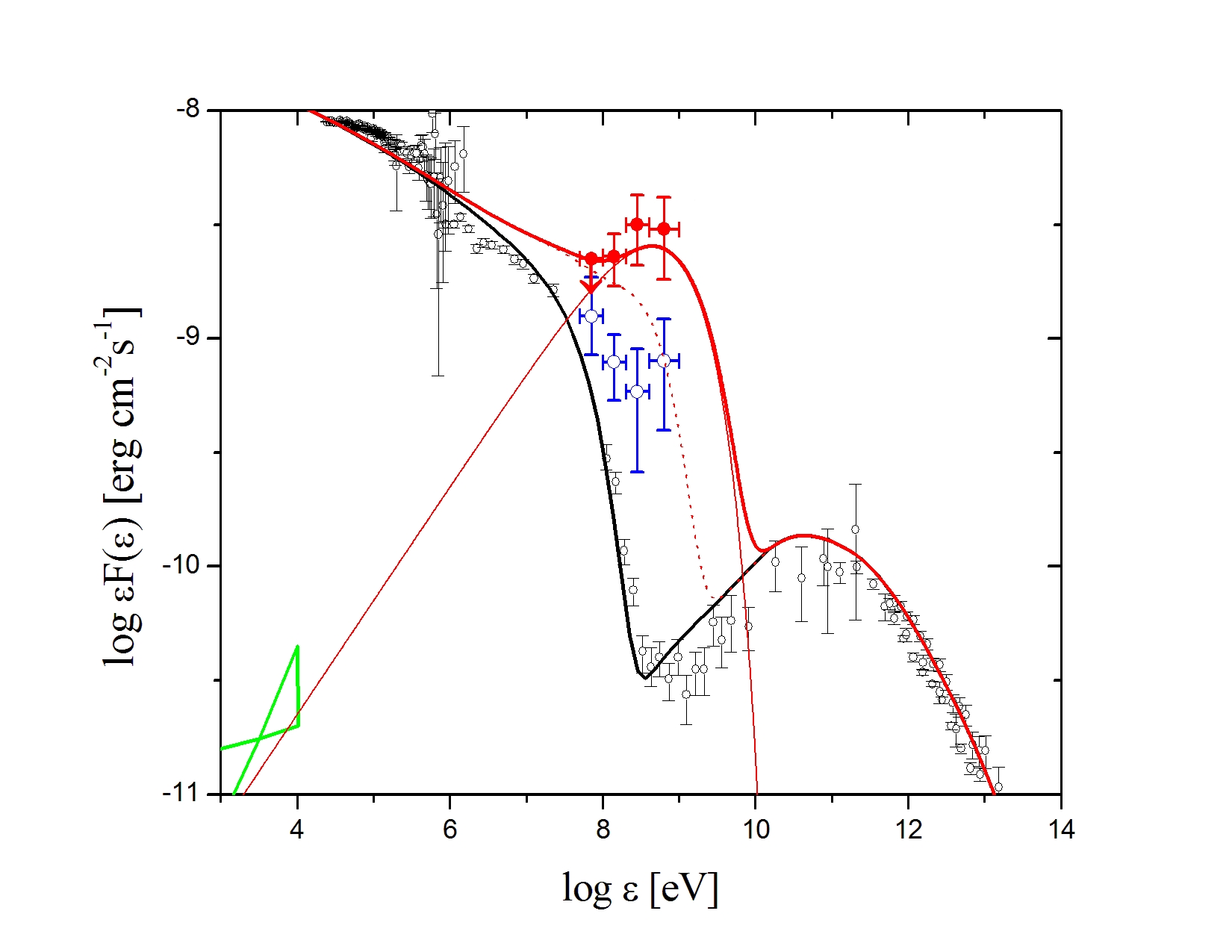}
\caption{AGILE gamma-ray (pulsar subtracted) spectra  of the Crab
Nebula for the April 2011 super-flare (red points, 1-day
integration) and for the September 2010 flare (blue points, 2-day
integration) (from ref. \cite{striani}). Black data points show
the standard Crab Nebula spectrum \cite{meyer}. The area marked in
green shows the X-ray emission spectrum detected by Chandra about
1 week after the Sept. 2010 flare from "Source A" in the inner
Nebula as reported in ref. \cite{tavani3}. }
 \label{fig-5}
 \end{center}
 \end{figure}


Very important information about the Crab inner nebula physical
processes can be gained from the  flaring spectral data.
Fig.~\ref{fig-4} shows the published 2-day averaged AGILE spectrum
and the 4-day \textit{Fermi}-LAT spectrum of the Sept. 2010 event.
Fig.~\ref{fig-5} shows the spectrum of the remarkable April 2011
record-high flare \cite{striani}. The Crab flare gamma-ray
emission above 100 MeV apparently violates Eqs.~\ref{eq-1} and
\ref{eq-2}, and we need to consider emission models different from
those applicable to the average nebular emission. Standard
theories based on MHD modelling with $E/B \leq 1$ and $\delta \sim
1$ can be applied to the average emission, but not to the flaring
states. This is a challenge for particle acceleration models that
in their standard forms are either too slow  (e.g., diffusive
models \cite{blandford,blandford2,drury}), not sufficiently
efficient \cite{kirk,reville}, or not directly applicable for the
probable lack of ions \cite{arons} in specific sites of the polar
jet regions such as the ''anvil". Relatively large scale MHD
instabilities (e.g., \cite{komissarov,komissarov2,delzanna,camus})
can play a role in setting the local conditions for fast
acceleration, but the current modelling does not capture the
flaring acceleration properties that most likely require a
detailed kinetic approach.
 Thus the Crab flaring data
may force us to substantially revise previously proposed models of
particle acceleration and emission . We notice here the  relevance
of the possible role of impulsive particle acceleration in
magnetic field reconnection, and/or runaway particle acceleration
by transient electric fields violating the condition $E/B < 1$.
The applicability of these concepts to the Crab Nebula flaring
activity remains to be tested by future investigations.

In order to model the observed flare spectra, we assume that a
sub-class of the overall particle population in the inner nebula
is subject to an impulsive acceleration process at a specific
site. In principle, favorably Doppler-boosted emission with
$\delta \sim 5-10$ might explain some properties of the gamma-ray
spectra of Figs.~\ref{fig-4} and \ref{fig-5}. However, it is not
clear whether this large Doppler boosting is supported by the Crab
Nebula data and geometry (for a discussion, see ref.
\cite{komissarov2}). Alternately, a particle acceleration
mechanism with and "effective" $E/B \ge 1$ and $\delta \sim 1$
seems to be required.
An example of theoretical modelling of the spectral evolution is
shown in Fig.~\ref{fig-4} for the Sept. 2011 event
\cite{vittorini}. Our modelling (aimed at testing the
compatibility for this event of the AGILE and \textit{Fermi}-LAT
spectra within a synchrotron cooling model) assumes an emitting
region of size $L = 7 \times 10^{15}\, $cm, and an enhanced local
magnetic field $B_{loc} = 10^{-3} \,$G.
%
 The acceleration process produces, within a timescale shorter
than any other relevant timescale,
 a particle energy distribution that we model as a double power-law
 distribution \cite{tavani3},
$ dn/d\gamma = K \, \gamma_b^{-1}/[(\gamma/\gamma_b)^{p_1}
+(\gamma/\gamma_b)^{p_2}] $,
where $n$ is the particle number density,  $p_1=2.1$, $p_2=2.7$,
$\gamma_b= 2 \times 10^9$, and $K$ a normalization factor. The
particle Lorentz factor $\gamma$ ranges from $\gamma_{min}=10^6$
and $\gamma_{max}=7\times 10^{9}$. This double power-law
distribution  implies maximal synchrotron emission between the
Lorentz factors $\gamma_b$ and $\gamma_{max}$.
The total particle number required to explain the flaring episode
turns out to be $ N_{e-/e+} = \int dV \, (dn/d\gamma) \, d\gamma =
2\times  10^{42}$, where $V$ is an assumed spherical volume of
radius $L$. Curves labelled in red color and in green color in
Fig.~\ref{fig-4} mark the results of the model calculation for
2-day and 4-day integrations, respectively. AGILE and
\textit{Fermi}-LAT data of the Sept. 2010 event, despite their
different integration timescales, are consistent with the rapid
synchrotron cooling model of ref. \cite{vittorini}.

Fig.~\ref{fig-5} shows the even more dramatic peak flaring
spectrum of the April 2011 event (from ref. \cite{striani}).
Spectral modelling of this event requires more extreme values of
the physical parameters compared to those applicable for the
September 2010 event. In particular, the hour-timescale
variability \cite{striani4,hays4} requires  very efficient
acceleration and fast synchrotron cooling that has to occur for
values of the local magnetic field $B_{loc} \ge 1$~mG. A strong
amplification of the local magnetic field is then a clear
signature of the highly transient acceleration process at work
(relativistic shocks, magnetic field reconnection, plasma
instabilities). Unveiling the fundamental process of fast
acceleration with extreme physical requirements is a challenging
opportunity for future theoretical work on the Crab.

\section{Chasing the acceleration site}

\begin{figure}
\begin{center}
\includegraphics[width=14cm]{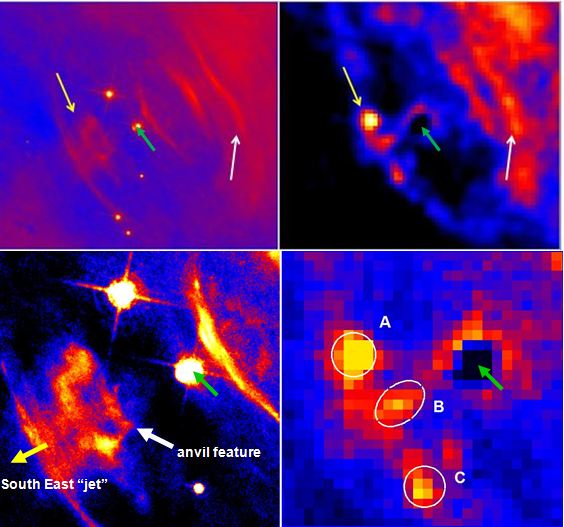}
\caption{ \textit{HST} and \textit{Chandra} imaging of the Crab
Nebula following the Sept., 2010 gamma-ray flare (from ref.
\cite{tavani3}). \textit{Top left panel:} optical image of the
inner nebula region (approximately 28"x28", North is up, East on
the left) obtained by the \textit{HST}-ACS instrument  on October
2, 2010. ACS bandpass: 3,500-11,000 Angstrom. The pulsar position
is marked with a green arrow in all panels. White arrows in all
panels mark interesting features compared to archival data.
\textit{Top right panel:} the same region imaged by the
\textit{Chandra} Observatory ACIS instrument on September 28, 2010
in the energy range 0.5-8 keV (level-1 data). The pulsar does not
show in this map and below because of pileup. \textit{Bottom  left
panel:} zoom of the \textit{HST} image (approximately 9"x9"),
showing the nebular inner region, and the details of the "anvil
feature" showing a "ring"-like structure at the base of the
South-East "jet" off  the pulsar. "Knot 1" at 0".6 South-East from
the pulsar is saturated at the pulsar position. Terminology is
from    ref. \cite{scargle}. \textit{Bottom right panel:} zoom of
the \textit{Chandra } image, showing the X-ray brightening of the
"anvil" region and the correspondence with the optical image.
Analysis of the features marked "A", "B", and "C" gives the
following results in the energy range 0.5-8 keV for the flux $F$,
spectral index $\alpha$, and absorption $N_H$ (quoted errors are
statistical at the 68\% c.l.). Feature A: flux $F = (48.5 \pm 8.7)
\cdot 10^{-12} \rm \, erg \,cm^{-2} \, s^{-1}$, $\alpha = 1.76 \pm
0.30$, $N_H = (0.36 \pm 0.05) \cdot 10^{22} \, \rm  atoms \,
cm^{-2}$. Feature B: flux $F = (26.6 \pm 5.9) \cdot 10^{-12} \rm
\,  erg \,  cm^{-2} \, s^{-1}$ , $\alpha = 1.76\pm 0.41$, $N_H =
(0.34 \pm 0.05) \cdot 10^{22} \rm \, atoms \,  cm^{-2}$. Feature
C: flux $F = (25.3 \pm 5.9) \cdot 10^{-12} \rm \,  erg \,  cm^{-2}
\, s^{-1}$, $\alpha = 1.46 \pm 0.36$ , $N_H = (0.34 \pm 0.04)
\cdot 10^{22} \, \rm atoms \, cm^{-2}$. }
 \label{fig-6}
 \end{center}
 \end{figure}

The range of the gamma-ray emission timescales (hours-days), and
the extreme energetics of the flaring phenomenon (reaching
gamma-ray luminosities  of $L_{\gamma,p} = (0.5-1) \times 10^{36}
\rm \, erg \; s^{-1}$ above 100 MeV)  indicate an emission site in
the inner Crab Nebula. Several regions can be considered for the
flaring particle acceleration site including: (1) the South-East
polar jet region and in particular the "anvil" site
\cite{tavani3,abdo2} (near  "knot-2" of
ref.~\cite{scargle,hester2}); (2) the inner ring (visible at the
optical and X-ray energies) marking the pulsar wind termination
shock; (3) the optical feature very close ($\sim 0".6$) to the
pulsar (possibly a standing shock) classified as "knot-1"
\cite{komissarov2}. High-resolution optical  and X-ray
observations of the inner Nebula are necessary to address this
issue. Fig.~\ref{fig-6} shows the results of \textit{Hubble Space
Telescope} optical and \textit{Chandra } X-ray observations of the
inner Crab Nebula following the Sept. 2010 gamma-ray flare
\cite{tavani3}. These observations were obtained 7 to 10 days
after the event, and show several active regions both in the
optical and in the X-ray energy bands. In particular, the "anvil"
region seems brightened compared to other historical observations
of the Nebula showing a less prominent emission. Several spots of
enhanced X-ray emission (especially those marked as "source-A" and
"source-B" in Fig.~\ref{fig-6}) can be noticed. Several local
transient enhancements of optical and X-ray emission have been
recorded in that region \cite{scargle,hester2,hester3,weisskopf}.
They typically last a few weeks, fade, and are  back in an
apparently random fashion after several months. It is tantalizing
to speculate that these features are associated with the fast
particle acceleration site producing the major Crab gamma-ray
flares, but the current data do not provide a "smoking gun" yet. A
program of X-ray monitoring of the Crab inner nebula by
\textit{Chandra} started in October 2010 envisioning regular
observations about 1 per month to provide an extensive baseline
\cite{weisskopf4}. \textit{Chandra} observed the Crab Nebula
several times in coincidence with the super-flare of April 2011
\cite{tennant}. These observations are currently being analyzed.

\section{Conclusions}

 The Crab Nebula continues to surprise us providing invaluable
 information on the most extreme acceleration processes.
 The pattern of well established gamma-ray flares shows a
 recurrence of about 1-2 events per year, leaving open the
 possibility that lower level gamma-ray enhanced emission can  be
 produced between major flares. The discovery of this
 remarkable phenomenon can be attributed to the monitoring
 capabilities of the current gamma-ray missions, AGILE and \textit{Fermi}.
Crab gamma-ray flares are not that rare, and we may wonder why
this phenomenon was not discovered earlier. The situation
regarding the overall variability in the Nebula is even more
interesting today after the recognition that the total X-ray flux
from the Crab is subject to long timescale variations of order of
years \cite{wilson}. It is not clear at present whether there is
any connection between the  pattern of year-long X-ray variations
\cite{qiang} and the fast and intense gamma-ray flaring. If the
central pulsar powerhouse determines a variable particle flux over
a long timescale, the high-energy activity we witness in the
Nebula can be ultimately traced back to the pulsar. Alternately,
plasma instabilities favored by large scale MHD flow variations
and/or magnetic field reconnection events can lead to the very
fast particle acceleration we require to explain the observations.
Future HST and Chandra observations of the inner Nebula will be
crucial to localize the particle acceleration site. We also note
that future TeV observations of the Crab will be important to test
Doppler-boosted models of emission \cite{striani}. A claim for
transient TeV emission in coincidence with the Sept. 2010 event
\cite{aielli} was not confirmed by independent short (20 min.)
observations \cite{ong,mariotti}. However, in light of the fast
variability recently determined in the Crab gamma-ray flares, this
issue needs to be addressed by future investigations
\cite{striani}. For sure, we can anticipate in the next years a
revival of theoretical and observational investigations of our
beloved Crab.

\section{Acknowledgements}

The material reported in this paper is based on the investigations
of the AGILE Team whose efforts and great work have been crucial
for the discovery of the dramatic gamma-ray variability of the
Crab Nebula. A special acknowledgement is for the excellent work
of the AGILE mission data processing and alert system that made
possible the discovery and the record-fast data analysis: we thank
M. Trifoglio, F. Gianotti, A. Bulgarelli, and the ASI Space Data
Center (ASDC) group. Specific work on the Crab data was performed
by E. Striani, V. Vittorini, I. Donnarumma, G. Piano, G. Pucella,
A. Trois, A. Pellizzoni, M. Pilia, Y. Evangelista, E. Del Monte,
C. Pittori, and F. Verrecchia. \textit{Chandra} observations of
the Crab Nebula were obtained and analyzed with the precious
collaboration of M. Weisskopf and A. Tennant. We thank P. Caraveo,
 A. De Luca, and R. Mignani for the collaboration and analysis of
\textit{HST} data. Research partially supported by the ASI grant
no. I/042/10/0.

\end{document}